\documentclass[aps,prd,twocolumn,nofootinbib,superscriptaddress]{revtex4}
\usepackage[hypertex]{hyperref}
\usepackage{graphicx}

\def\figsize{3 in}

\newcommand{\be}{\begin{equation}}  
\newcommand{\ee}{\end{equation}}  
\newcommand{\bea}{\begin{eqnarray}}     
\newcommand{\eea}{\end{eqnarray}}     

\begin{document}

\pagestyle{plain}

\title{A Composite Twin Higgs Model}

\author{Puneet Batra} \affiliation{Department of Physics, Columbia
University, 538 W. 120th St., NYC, NY 10027} \author{Z. Chacko}
\affiliation{Department of Physics, University of Maryland, 
College Park, MD, 20742}
   

\begin{abstract}

Twin Higgs models are economical extensions of the Standard Model that stabilize the 
electroweak scale. In these theories the Higgs field is a pseudo Nambu-Goldstone boson that 
is protected against radiative corrections up to scales of order 5 TeV by a discrete parity 
symmetry.  We construct, for the first time, a class of composite twin Higgs models based on 
confining QCD-like dynamics. These theories naturally incoporate a custodial isospin symmetry 
and predict a rich spectrum of particles with masses of order a TeV that will be accessible 
at the LHC.


\end{abstract}

 \maketitle


\section{Introduction}

Quantum corrections to the Higgs mass parameter in the Standard Model are quadratically 
divergent. Stabilizing the weak-scale against these divergences generally requires a 
phenomenologically rich spectrum of new particles near a TeV, associated with the existence 
of a new symmetry of nature. One appealing idea is that the Higgs sector of the Standard 
Model (SM) might actually be the non-linear sigma model of some larger, dynamically generated 
pattern of symmetry breaking \cite{GP, KG}. In such theories, the Higgs behaves much like a 
pion in QCD; the Higgs is a composite pseudo Nambu-Goldstone boson (pNGB) which is protected 
from the worst class of quadratic divergences that usually afflict scalar bosons. Precision 
electroweak measurements currently constrain the compositeness scale to lie above 5 TeV. The 
fact that the SM gauge couplings, top Yukawa coupling and Higgs self-coupling necessarily 
break any global symmetry with order one strength then implies that the mass parameter of the 
composite Higgs needs additional protection, if the theory is to be natural.

Little Higgs theories~\cite{Little1},~\cite{Little2},~\cite{NMM}, for reviews 
see~{\cite{Reviews}} are a class of non-linear sigma models which realize the Higgs as a 
protected pNGB. The underlying concept behind little Higgs theories is the idea of 
`collective symmetry breaking' - the global symmetry of which the Higgs is the pNGB is broken 
only when two or more couplings in the Lagrangian are non-vanishing.  This is a significant 
restriction on the form of the quantum corrections to the pNGB potential, which can be used 
to engineer natural electroweak symmetry breaking. These theories stabilize the weak scale to 
about 5-10 TeV.

Twin Higgs theories~\cite{twin},~\cite{twinLR} are an alternative class of non-linear sigma 
models which also realize the Higgs as a protected pNGB. These theories possess a discrete 
$Z_2$ interchange symmetry, in addition to the approximate global symmetry of which the Higgs 
is the pNGB. In the existing twin Higgs models this $Z_2$ symmetry is identified either with 
mirror symmetry, or with left-right symmetry. This discrete symmetry is enough to ensure that 
any quadratically divergent contribution to the scalar potential accidentally respects the 
global symmetry, and therefore cannot contribute to the mass of the pNGB. These theories also 
stabilize the weak scale to about 5-10 TeV.

Since in general little Higgs and twin Higgs theories have been formulated only as non-linear 
sigma models, above 5-10 TeV these theories require ultra-violet completions to maintain 
unitarity. Weakly coupled ultra-violet completions using supersymmetry have been constructed 
for both the little Higgs~\cite{littleSUSYold},~\cite{littleSUSY} and the twin 
Higgs~\cite{twinSUSY}, in the context of the supersymmetric little hierarchy problem. In the 
little Higgs case, non-supersymmetric ultra-violet completions have also been 
constructed~\cite{turtle},~\cite{tower}~\cite{CHPS}. There has also been some work on the 
difficult problem of realizing the little Higgs as a strongly coupled 
composite~{\cite{KLNW},~\cite{Thaler:2005kr},~{\cite{Hill}},~\cite{BC1},~\cite{holographic}} 
furthering the analogy to QCD. However, in the twin Higgs case the corresponding problem has 
not been addressed.

A significant challenge in dynamically realizing a composite twin Higgs is to ensure that the 
strong dynamics respects a custodial SU(2) symmetry. For this to happen, the custodial 
symmetry must be contained in the non-linearly realized global symmetry of the Higgs sector. 
In the twin Higgs models currently in the literature this global symmetry is either SU(4), 
which is spontaneously broken to SU(3), or O(8), spontaneously broken to O(7). While the 
breaking of SU(4) to SU(3) is fairly straightforward to realize through QCD-like strong 
dynamics~{\cite{Thaler:2005kr}}, this pattern does not admit a custodial SU(2). On the other 
hand, while the O(8) $\rightarrow$ O(7) pattern does preserve a custodial symmetry, this 
pattern is significantly more complicated to realize through strong dynamics.

In this paper we identify an alternative pattern of symmetry breaking for twin Higgs models 
which naturally incorporates a custodial isospin symmetry. We then show how this pattern can 
be realized through QCD-like dynamics, and apply these ideas to construct a class of 
composite twin Higgs models with left-right symmetry.  These theories predict a rich spectrum 
of new particles at the TeV scale that will be accessible to the LHC.

We begin by constructing an alternative realization of the twin Higgs model, in its 
left-right symmetric incarnation. Consider a scalar field $H$ which transforms as a 
fundamental under an Sp(4) global symmetry, and which is also charged under a global U(1). If 
$H$ acquires a VEV such that $\langle H \rangle = (0,0,0,f)$, the Sp(4)$\times$U(1) global 
symmetry is spontaneously broken to SU(2)$\times$U(1), and there are 7 Goldstone bosons. We 
now break the global Sp(4) explicitly by gauging an SU(2$)_{\rm L}\times$SU(2$)_{\rm R}$ 
subgroup. The overall U(1), which is to be identified with U(1$)_{{\rm B}-{\rm L}}$, is also 
gauged. The overall gauge structure is therefore that of a left-right symmetric 
model~{\cite{LR}}.

Under gauge transformations the field $H$ decomposes into $\left({ H_L , H_R} \right)$ where 
$H_L$ is a doublet under SU(2$)_{\rm L}$ and $H_R$ as a doublet under SU(2$)_{\rm R}$. If the 
VEV of $\langle H \rangle$ points along a direction which breaks SU(2$)_{\rm R}$ but 
preserves SU(2$)_{\rm L}$, the surviving gauge symmetry is the familiar SU(2$)_{\rm 
L}$$\times$U(1$)_{\rm Y}$ of the SM. Of the 7 Goldstone bosons, 3 are eaten. The remaining 4 
Goldstone bosons, which are contained in $H_L$, are to be identified with the SM Higgs. If 
the discrete parity symmetry, which interchanges SU(2$)_{\rm L}$ and SU(2$)_{\rm R}$, is 
exact, $H_L$ is protected against quadratic divergences by the twin Higgs mechanism. The key 
observation is that the discrete symmetry ensures that any quadratically divergent 
contribution to the scalar potential has an Sp(4) invariant form, and therefore cannot 
contribute to the mass of the Goldstones.

Yukawa interactions can take the same form as in the original left-right twin Higgs model, 
since they are only required to respect the gauge and parity symmetries, which are identical 
in both models.  Although these couplings violate the global symmetry with order one 
strength, the discrete parity symmetry again ensures that quadratic divergences are absent. 
From this we infer that [Sp(4)$\times$U(1)]/[SU(2)$\times$U(1)] constitutes an alternative 
symmetry breaking pattern which allows the realization of a twin Higgs model with left-right 
symmetry.

Although this construction is extremely simple, it does not admit a custodial SU(2) symmetry. 
Furthermore, it is not clear whether such a pattern of symmetry breaking can arise from 
strong dynamics. In the next section we show that a natural generalization of this model 
exists which addresses the first problem. We then go on to discuss how the required symmetry 
breaking pattern can be realized through the condensation of strongly coupled fermions, in 
analogy with QCD.

\section{A Custodial Symmetry for the Twin Higgs} 

Consider a theory with an Sp(4) $\!\times\!$ Sp(4) global symmetry, which is 
spontaneously broken down to the diagonal Sp(4). We label the 10 resulting 
NGBs that are produced by $\pi^A$, and define
\begin{equation}
 X = f {\rm exp}\left({2 i}\pi^A T^A/f \right). 
\end{equation}
Here the matrices $T^A$ are the generators of Sp(4), and correspond to 
the matrices
\begin{equation} 
\pmatrix{ \sigma^a & 0 \cr
          0 & 0}
\pmatrix{ 0 & 0 \cr
          0 & \sigma^a}
\pmatrix{ 0 & iI \cr
         -iI & 0}
\pmatrix{ 0 & \sigma^a \cr
         \sigma^a & 0}.
\label{sp4gen}
\end{equation}
We now gauge an SU(2$)_{\rm L}$$\times$SU(2$)_{\rm R}$ subgroup of the 
first Sp(4), and an SU(2$)_{\rm L'}$$\times$U(1$)_{\rm R'}$ subgroup of 
the second Sp(4). Here U(1$)_{\rm R'}$ is the diagonal generator of the 
SU(2$)_{\rm R'}$ contained in the second Sp(4). We label the gauge 
coupling constants of these four groups as $g_L$, $g_R$, $g'_L$ and $g'_R$ 
respectively. The unbroken gauge symmetry is then SU(2)$\times$U(1), 
which is identified with the electroweak gauge group of the SM. Note that 
this symmetry breaking pattern is similar to that of the little Higgs 
model of Chang and Wacker~\cite{NMM}. Of the original 10 Goldstone bosons, 
6 are eaten. The remaining 4 pseudo-Goldstone bosons are identified with 
the SM Higgs, and correspond to the generators $T^a$,
\begin{equation}
\left\{ T^a \right\} =  \left\{
\pmatrix{ 0 & iI \cr
         -iI & 0}
\pmatrix{ 0 & \sigma^a \cr
         \sigma^a & 0}
\right\}.
\label{unbrokensp4gen}
\end{equation}

We can write an effective field theory for the pNGBs which is valid at low momenta. This 
takes the form of a non-linear sigma model. In general the Lagrangian for this theory will 
contain all operators involving the field $X$ consistent with the non-linearly realized Sp(4) 
$\times$ Sp(4) global symmetry. Non-renormalizable operators are suppressed by the cutoff 
$\Lambda$ of the non-linear sigma model, and their coefficients are determined by the 
specific ultra-violet completion. The cutoff $\Lambda$ must be less than about $4 \pi f$, 
where the upper bound corresponds to strong coupling.

In this low-energy theory, the masses of the pseudo-Goldstones are protected against one loop 
quadratic divergences from gauge interactions. This can be understood as a consequence of the 
little Higgs mechanism. The theory has an exact Sp(4) global symmetry in the limit that $g_L$ 
and $g_R$ are zero, and also in the limit that $g_L'$ and $g_R'$ are zero. Any diagram that 
results in a quadratic divergence must therefore involve both these sets of couplings. The
leading contributions to the pseudo-Goldstone masses arise at order $g^4$, and are therefore
necessarily suppressed by at least two loop factors. 

If the low-energy effective theory is weakly coupled at the scale $\Lambda$, in the special 
case that the SU(2$)_{\rm L}$ and SU(2$)_{\rm R}$ of the first Sp(4) are related by a 
discrete interchange symmetry, so that $g_L = g_R$, there is an alternative way of 
understanding this cancellation based on the twin Higgs mechanism. This discrete symmetry 
ensures that at quadratic order in $X$ all radiative corrections to the pseudo-Goldstone 
potential are invariant under the first Sp(4), and therefore must simply vanish. To see this 
let us consider all possible operators consistent with the gauge symmetry at quadratic order 
in $X$ in the non-linear sigma model. At one loop these terms are the only ones generated 
with a quadratically divergent coefficient in the effective potential. Schematically these 
operators include
 \begin{eqnarray} 
X_{LL'} X^{\dagger \; L'L} && 
X_{RL'} X^{\dagger \; L'R} \nonumber \\ 
X_{L3} X^{\dagger \; 3L} && X_{R3} X^{\dagger \; 3R} \nonumber \\ 
X_{L4} X^{\dagger \; 4L} && X_{R4} X^{\dagger \; 4R} 
 \end{eqnarray} 
and also (suppressing hermitian conjugates) 
 \begin{eqnarray} 
\epsilon_{LL} \epsilon_{L'L'} X_{LL'}X_{LL'} && 
\epsilon_{RR} \epsilon_{L'L'} X_{RL'}X_{RL'} \nonumber \\ \epsilon_{LL} X_{L3}X_{L4} && 
\epsilon_{RR} X_{R3}X_{R4} 
 \end{eqnarray} 
Here $L$ and $L'$ take values 1 and 2, while $R$ takes values 3 and 4.  The discrete $L 
\leftrightarrow R$ symmetry ensures that in the Lagrangian operators on the same line above 
necessarily have the same coefficient. Then it is clear that at quadratic order in $X$ the 
Lagrangian is actually invariant under the first global Sp(4) symmetry. This symmetry is only 
broken at quartic order in $X$, and therefore corrections to the pNGB mass are loop 
suppressed, and at most logarithmically divergent.

The argument above does not carry over to the case where the low-energy theory is strongly 
coupled at the cutoff $\Lambda$, because now the quartic terms in $X$, though still loop 
suppressed, need not be small. The reason is that the quartic terms can now be generated at 
order $g^2$ by loops involving operators which are strongly coupled at the cutoff, and this 
could potentially compensate for the loop suppression.\footnote{Whether a quartic term is 
generated at order $g^2$ in a general twin Higgs model in the limit of strong coupling 
depends on the pattern of symmetry breaking. For example, if the symmetry breaking pattern is 
O(8) $\rightarrow$ O(7), a quartic is only generated at order $g^4$, and is therefore always 
loop suppressed~{\cite{twinLR}.}} However, the little Higgs mechanism still ensures that any 
such term is invariant under the second Sp(4), and so does not contribute to the pNGB 
potential. Therefore the leading terms which contribute to the mass of the pNGB only arise at 
order $g^2 g'^2$, and are suppressed by an additional loop factor.

This construction ensures that the strong dynamics does not violate the custodial 
SU(2) symmetry.  To see this explicitly, note that we can write
\bea
2\pi^a T^a &\equiv &  \pmatrix{ 0 & \phi \cr \phi^{\dagger} & 0}
\eea 
where $\phi = \left( i \sigma_2 {h}_L^*, {h}_L \right)$.  The 
full expression for $X$ is
\begin{equation}
 \cos \left( {|{h}_L|\over f} \right){f}   \nonumber \\
+ {if \over |{h}_L|}  \sin \left( {|{h}_L| \over f}  \right)
\pmatrix{ 0 & \phi \cr
         \phi^{\dagger} & 0}
\end{equation}
The Lagrangian written as a function of $X$ preserves the SU(2$)_{\rm L}$$\times$SU(2$)_{\rm R}$ subgroup of the diagonal Sp(4), under which 
$\phi \rightarrow U_L \phi U_R^{\dagger}$, and $|{h}_L| \rightarrow |{h}_L|$. After 
electroweak symmetry breaking, $\langle {h}_L \rangle = (0, v)$, and the diagonal SU(2) symmetry is preserved. This 
is precisely the custodial symmetry we are looking for.

In order to write down Yukawa couplings, first make the 
identification
 \begin{equation}
X_{i4} = {H}_i = ({H}_L, {H}_R).
 \end{equation}
Yukawa couplings can be written down exactly as in the original left-right twin Higgs model, 
in terms of ${H}_L$ and ${H}_R$, so that the discrete $L \leftrightarrow R$ symmetry is 
preserved. The twin Higgs mechanism then ensures that quadratic divergences 
from the fermion sector preserve the first global Sp(4) symmetry and vanish from the 
pseudo-Goldstone potential, just as in the gauge sector.
 
The fermionic content of the theory then contains three generations of
\begin{eqnarray}
Q_L = \left(u,d \right)_L =\left[2,1,1/3 \right] \; \;
L_L = \left(\nu,e \right)_L = \left[2,1, -1 \right] \nonumber \\
Q_R = \left(u,d \right)_R = \left[1,2, 1/3 \right] \; \;
L_R = \left(\nu,e \right)_R = \left[1,2, -1 \right]
\end{eqnarray}
where the square brackets indicate the quantum numbers of the corresponding field 
under SU(2$)_{\rm L}$$\times$SU(2$)_{\rm R}$$\times$ U(1$)_{\rm B - L}$.  We 
identify U(1$)_{\rm R'}$ with U(1$)_{\rm (B - L)/2}$. As dictated by left-right 
symmetry the theory includes right-handed neutrinos in addition to the SM 
fermions.

The Higgs fields have quantum numbers
\begin{equation}
{H}_L = \left[2, 1, 1
\right] \; \; \; \; \; \; {H}_R = \left[1,2,1 \right]
\end{equation}
under SU(2$)_{\rm L}$$\times$SU(2$)_{\rm R}$$\times$U(1$)_{\rm B - L}$.
The down-type Yukawa couplings of the SM arise from non-renormalizable
couplings of the form
\begin{equation}
\left\{\frac{\overline{Q}_R {H}_R {H}_L^{\dagger} Q_L \; \; + \;\;
\overline{L}_R {H}_R {H}_L^{\dagger} L_L}{\Lambda} \right\}  \; \; + \; \;   
{\rm h.c.}
\end{equation}
Here $\Lambda$ is an ultra-violet cutoff, which we take to be about 10 
TeV, the limit of validity of the non-linear sigma model. 
Similarly, the up-type Yukawa couplings of the SM emerge from
\begin{equation}
\left\{\frac{\overline{Q_R} \; {H}_R^{\dagger} {H}_L Q_L \; \; + \;\;
{\rm h.c.}}{\Lambda} \right\}
\end{equation}

The top Yukawa coupling is too large to be naturally obtained from a 
non-renormalizable operator. As in the original left-right twin Higgs model, we 
therefore introduce a pair of vector-like quarks $T_L$ and $T_R$ which have the 
quantum numbers
\begin{equation}
T_L = \left[1, 1, 4/3  \right] \; \; \; \; \; \;
T_R = \left[1, 1, 4/3  \right]
\end{equation}
under SU(2$)_{\rm L}$$\times$SU(2$)_{\rm R}$$\times$U(1$)_{\rm B - L}$. We
can then write
the Yukawa coupling
\begin{equation}
\label{topmodule}
\left( y \; \overline{Q_R} \; {H}_R^{\dagger} T_L \; + \;
y \; \overline{Q_L} \; {H}_L^{\dagger} T_R \; + \;
M \overline{T_L} T_R \right)  \; \; + \; \; {\rm h.c.}
\end{equation}
Here $Q_L$ and $Q_R$ are the usual left and right-handed third
generation quark doublets of the left-right model.

Since the top Yukawa gives the largest contribution to the Higgs potential, let us understand 
the cancellation of quadratic divergences in this case. As in the gauge case, the discrete $L 
\leftrightarrow R$ symmetry ensures that terms quadratic in $X$ are invariant under the first 
Sp(4), and do not contribute to the potential for the pNGB. Terms quartic and higher order in 
$X$ that violate Sp(4) are only generated at order $y^4$, and not at order $y^2$, and are 
therefore suppressed by one loop factor, even in the limit that the non-linear sigma model 
is strongly coupled at the cutoff.

It is also possible to generate the smaller Yukawa couplings from 
renormalizable interactions~\cite{ALRM}, (see also~\cite{RabiLR}). To do 
this we introduce three generations of vector-like fermions with the 
following charge assignments.
\begin{eqnarray}
U_L = \left[1, 1, 4/3  \right] \; \; \; \; \; \;
U_R = \left[1, 1, 4/3  \right] \nonumber \\
D_L = \left[1, 1, -2/3  \right] \; \; \; \; \; \;
D_R = \left[1, 1, -2/3  \right] \nonumber \\
E_L = \left[1, 1, -2  \right] \; \; \; \; \; \;
E_R = \left[1, 1, -2  \right]
\end{eqnarray}
Then the Yukawa couplings for the lighter fermions can be written down in 
analogy with that for the top. For example, the charged lepton Yukawa 
couplings arise from the interactions 
\begin{equation}
\left\{ \overline{L_R} \; {H}_R E_L \; + \;
\overline{L_L} \; {H}_L E_R \; + \;
M \overline{E_L} E_R \right\}  \; \; + \; \; {\rm h.c.}
\end{equation}
We choose the mass parameter $M$ to be of order several TeV. On 
integrating out $E_L$ and $E_R$ we get back exactly the same 
non-renormalizable operator that earlier generated the charged lepton 
masses.

\section{A Twin Higgs Model from Strong Dynamics} 

We now explain how the symmetry breaking pattern Sp(4)$\times$Sp(4) 
$\rightarrow$ Sp(4) may be obtained from QCD-like strong dynamics. Our 
discussion will closely follow that of~\cite{BC1}, where the same problem 
was considered in the context of a dynamical realization of the 
little Higgs model of Chang and Wacker~\cite{NMM}. Consider an SU(${\rm 
N_c}$) gauge 
group, with a set of four fermions, $\chi_{\alpha i}$, in the fundamental 
representation. Here $\alpha$ represents an SU(${\rm N_c}$) gauge index 
and $i$ labels the fermions from 1 through 4. We also add a set of four 
right-handed fermions  ${\psi}_{\alpha i}$. When the 
SU(${\rm N_c}$) theory gets strong, a condensate $\langle 
\chi_{i} \overline{\psi}_j \rangle \propto \delta_{ij}$ forms 
and breaks the SU(4$)^2$ flavor symmetry to the diagonal SU(4). We 
label the 15 resulting NGBs that are produced by $\pi^A$, and define $X = 
f {\rm exp}\left({2 i}\pi^A T^A/f \right)$, where the matrices $T^A$ are 
generators of SU(4). We also add to the theory a non-renormalizable term 
\begin{equation}
\label{eq:explicit} 
\frac{m^2}{\left(4 \pi f^2 \right)^2} {\rm Tr} \left[ 
\left( \chi\overline{\psi} \right) J \left( \chi \overline{\psi} \right)^T J \right] 
\sim m^2 {\rm Tr} \left[ X J X^T J \right] 
\end{equation} 
where $J$ is the matrix 
\begin{equation} 
J = \pmatrix{ i \sigma^2 & 0\cr
                         0& i \sigma^2}.
\end{equation}
The effect of this term is to explicitly break the global ${\rm SU(4)}^2$ 
symmetry to ${\rm Sp(4)}^2$, thereby giving a mass of order $m$ to 5 of 
the 15 NGBs. With the addition of this term the pattern of global symmetry 
breaking is in fact ${\rm Sp(4)^2 \! \rightarrow \! Sp(4)}$, which 
accounts for the 10 surviving NGBs. The unbroken global symmetry, the 
diagonal Sp(4), contains the custodial SU(2) symmetry we desire.

\begin{figure}[t] 
\includegraphics[width=\figsize]{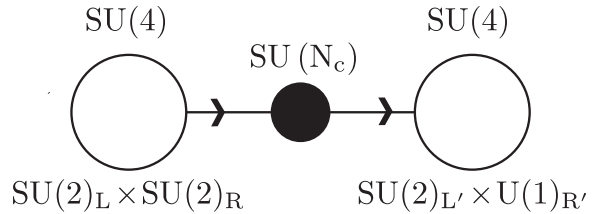} 
\caption{A UV completion for the custodial Twin Higgs model. The theory has a global 
SU(4$)^2$ flavor symmetry, with the indicated gauged subgroups. The two link fields represent 
the sets of SU(N$_c$) fundamentals, $\psi_i$ and $\chi_j$.}
\label{fig:uv} 
\end{figure}

In order to recreate the low energy structure of the model of the previous 
section we gauge the subgroups \be \left[{\rm SU(2)_L} \! \times \! {\rm 
SU(2)_R}\right] \! \times \! \left[{\rm SU(2)_{L'}} \! \times \! {\rm 
U(1)_{R'}}\right] \subset {\rm Sp(4)}^2 \ee as shown in Figure \ref{fig:uv}. 
After symmetry breaking, this gauge symmetry is broken down to the ${\rm 
SU(2)} \! \times \! {\rm U(1)_Y}$ gauge symmetry of the SM, where ${\rm 
SU(2)}$ is the diagonal subgroup of ${\rm SU(2)_L} \! \times \! {\rm 
SU(2)_{L'}}$, while ${\rm U(1)_Y}$ is the unbroken linear combination of 
the diagonal generator of ${\rm SU(2)_{R}}$ and ${\rm U(1)_{R'}}$. Of the 
10 surviving NGBs, 6 are eaten by the broken gauge symmetries, while the 4 
which are left over precisely constitute the SM Higgs.

In order to write down the Higgs couplings to fermions we simply
make the replacement
\begin{equation}
{H}_i \rightarrow \frac{ {\chi}_{i}\overline{\psi}_{4}}{ 4 \pi f^2}
\end{equation}
in the Yukawa couplings of the previous section. For example, the
left-right symmetric top Yukawa couplings become
\begin{equation}
\left\{ y \; \overline{Q_R} \; 
\left(\frac{\chi_R \overline{\psi}_{4} } { 4 \pi f^2} \right) T_L 
\; + \; y \; \overline{Q_L} \; 
\left(\frac{\chi_L \overline{\psi}_{4}}{ 4 \pi 
f^2}\right)^{\dagger} T_R \right\}
\end{equation}
These interactions are non-renormalizable, and therefore require 
additional new physics to generate them. We leave the question of the 
ultra-violet origin of these operators for future work.

We briefly consider the precision electroweak constraints on this theory. In general, bounds 
from the S-parameter on any composite Higgs force the compositeness scale $\Lambda \sim 4 \pi 
f$ to be larger than or of order 5 TeV. Another source of corrections to precision 
electroweak observables arises from higher order operators in the expansion of the kinetic 
term for $X$ in terms of the $\pi$ fields that contribute to the $\rho$ parameter.  Since the 
sum over the $T^A$ in $X = f {\rm exp}\left({2 i}\pi^A T^A/f \right)$ now runs over all SU(4) 
generators, and not just the generators of Sp(4), there are fields with mass below the 
compositeness scale that correspond to the generators of SU(4)/Sp(4). These fields, which we 
denote by $H'_L$, have exactly the same gauge quantum numbers as the light Higgs. The 
non-renormalizable terms that arise in the expansion of $X$ in terms of the $\pi$ fields 
involve custodial SU(2) violating couplings of $H'_L$ to the light Higgs $H_L$, and thereby 
contribute to the $\rho$ parameter.

The effect of the non-renormalizable term in Eq.~({\ref{eq:explicit}}) is to give a mass $m$ 
to the fields in $H'_L$, and to thereby decouple them from the low-energy spectrum. The 
precision electroweak constraints therefore translate into a lower bound on the parameter 
$m$. A quick estimate of the size of the correction to $\rho$ yields
\begin{equation}
\frac{\delta m^2_Z}{m^2_Z} \sim \frac{\langle H'_L \rangle^2}{f^2} 
\end{equation}
A VEV for $H'_L$ arises from the radiatively generated mass term which mixes 
$H'_L$ with the light Higgs.
\begin{equation}
\langle H'_L \rangle \sim \left( \frac{f}{4 \pi m} \right)^2 v
\end{equation}
Here $v$ is the electroweak VEV. From these formulas we estimate that the precision 
electroweak constraints on deviations of the $\rho$ parameter from its SM value are 
comfortably satisfied provided that $m$ is greater than or of order 500 GeV.

As in any general two Higgs doublet model, the presence of a second Higgs doublet can also 
lead to contributions to the $\rho$ parameter at loop level. However, in this specific model, 
this contribution translates into a lower bound on $m$ somewhat weaker than the one we have 
already found.

The additional SU(2$)_{\rm R}$, U(1$)_{\rm B - L}$ and SU(2$)_{\rm L'}$ gauge bosons also 
contribute to the precision electroweak observables ~\cite{NMM}. In general these force $f$ to be of 
order 1500 GeV or larger, reintroducing fine-tuning. However, $f$ can be as low as 500 GeV 
if, as in the original left-right twin Higgs model, there is a second field $\hat{X}$ with 
exactly the same quantum numbers as $X$ that exhibits exactly the same pattern of symmetry 
breaking, but where the decay constant $\hat{f}$ somewhat larger than $f$. Then, provided 
that $\hat{f}$ is greater than about 1500 GeV the precision electroweak constraints from the 
new gauge bosons are satisfied, and the fine-tuning is under control. The field $\hat{X}$ can 
also be used to generate neutrino masses~\cite{twinneutrino} and dark 
matter~\cite{twindarkmatter}, as in the original left-right twin Higgs model.
  
Much of the heavy spectrum of particles predicted by this theory will be accessible at the LHC. The new fields in the gauge and top sector 
must have mass of order the TeV scale if they are to be relevant for stabilizing the electroweak scale. Production and decay of the heavy 
top-partner, as well as the massive gauge bosons associated with SU(2$)_{\rm R}$ and U(1$)_{\rm B - L}$, have been studied in the context of 
the left-right twin Higgs model \cite{Goh:2006wj}. The heavy electroweak doublet of scalars, from the explicit breaking of the global ${\rm 
SU(4)}^2$ symmetry to ${\rm Sp(4)}^2$ in Eq.~(\ref{eq:explicit}), and the massive gauge bosons that constitute the linear combination of 
SU(2$)_{\rm L}$ and SU(2$)_{\rm L'}$ that is orthogonal to SU(2) of the SM, are key predictions of the underlying composite structure of this 
model.  While the electroweak doublet decays primarily into third generation quarks and anti-quarks, the new gauge bosons can decay either 
into SM fermions, or into electroweak gauge bosons.

In summary we have identified a new class of left-right twin Higgs models which naturally 
incorporate a custodial SU(2) symmetry, and shown how the relevant pattern of symmetry 
breaking can be realized through QCD-like strong dynamics. This constitutes an important 
first step in the construction of completely realistic twin Higgs models from strong 
dynamics.

\begin{acknowledgments}
PB is supported by the DOE under contract DE-FG02-92ER. ZC is supported by the NSF under 
grant PHY-0801323.
\end{acknowledgments}

\end{document}